\documentclass[10pt,aps,twocolumn,pra,superscriptaddress,showpacs,longbibliography]{revtex4-1}

\usepackage{graphicx}
\usepackage{dcolumn}
\usepackage{bm}
\usepackage{upgreek}
\usepackage{paralist}
\usepackage{ifthen}
\usepackage[english]{babel}

\begin{document}

\newcommand{\EECS}{\affiliation{Department of Electrical Engineering and Computer Science, Massachusetts Institute of Technology, Cambridge MA 02139, United States}}
\newcommand{\Columbia}{\affiliation{Department of Electrical Engineering, Columbia University, New York, New York 10027, United States}}

\title{Review Article: Quantum Nanophotonics in Diamond}

\author{Tim Schr\"{o}der}
\EECS
\author{Sara Mouradian}
\EECS 
\author{Jiabao Zheng} 
\EECS
\Columbia
\author{Matthew E. Trusheim}
\EECS
\author{Michael Walsh}
\EECS
\author{Edward H. Chen}
\EECS
\author{Luozhou Li}
\EECS
\author{Igal Bayn}
\EECS
\author{Dirk Englund}
\EECS
\date{\today}

\begin{abstract}
The past decade has seen great advances in developing color centers in diamond for sensing, quantum information processing, and tests of quantum foundations. Increasingly, the success of these applications as well as fundamental investigations of light-matter interaction depend on improved control of optical interactions with color centers -- from better fluorescence collection to efficient and precise coupling with confined single optical modes. Wide ranging research efforts have been undertaken to address these demands through advanced nanofabrication of diamond.  This review will cover recent advances in diamond nano- and microphotonic structures  for efficient light collection, color center to nanocavity coupling, hybrid integration of diamond devices with other material systems, and the wide range of fabrication methods that have enabled these complex photonic diamond systems.
\end{abstract}

\maketitle
\tableofcontents

\section{Introduction}
\label{sec:Intro}
Over the past two decades, color centers in diamond have emerged as promising systems for quantum information (QI) applications~\cite{wrachtrup_processing_2006,childress_diamond_2013,nemoto_photonic_2013} and precision sensing~\cite{wrachtrup_nitrogen-vacancy_2013,hong_nanoscale_2013,schirhagl_nitrogen-vacancy_2014,rondin_magnetometry_2014}. They were the first and are still among the brightest solid-state, room-temperature single photon sources~\cite{brouri_photon_2000,kurtsiefer_stable_2000}. Moreover, several defects allow for optical access to associated electron and nuclear spin states, which can exhibit long coherence times~\cite{bar-gill_solid-state_2013}, enabling their use as quantum memories~\cite{maurer_room-temperature_2012} in QI applications~\cite{gao_coherent_2015}. These defects also exhibit strong sensitivity to magnetic field~\cite{maze_nanoscale_2008,hong_nanoscale_2013}, electric field~\cite{dolde_electric-field_2011}, strain~\cite{ovartchaiyapong_dynamic_2014}, pressure~\cite{cai_signal_2014}, and temperature dependence~\cite{neumann_high-precision_2013}, enabling sensing of small fields at low frequencies~\cite{clevenson_broadband_2015,wolf_subpicotesla_2015}, often at room temperature~\cite{le_sage_efficient_2012}, and down to the single nuclear spin level~\cite{taylor_high-sensitivity_2008,lovchinsky_nuclear_2016}. 
By coupling these already promising defect centers to optical nano- and microstructures~\cite{babinec_diamond_2010,hadden_strongly_2010,faraon_resonant_2011,riedrich-moller_one-_2012,li_coherent_2015}, one can shape and control the optical properties to increase the performance, efficiency, and fidelity of sensing and QI protocols. 

Recent demonstrations of a variety of diamond patterning techniques -- focused ion beam (FIB) milling~\cite{bayn_triangular_2011}, reactive ion etching (RIE)~\cite{ando_smooth_2002,hausmann_fabrication_2010,li_nanofabrication_2015}, quasi-isotropic etching~\cite{khanaliloo_high-q/v_2015}, and electron beam induced etching (EBIE)~\cite{martin_subtractive_2014} -- have enabled patterning of diamond at the nanoscale, and thus the field of diamond nanophotonics. With high quality diamond fabrication now a reality, many optical systems have been proposed and realized in diamond and hybrid diamond systems. Here, we will discuss photonic structures for increased collection efficiency, stand-alone defect-cavity systems for tailored light-matter interaction, and hybrid photonic architectures for photon collection, routing, interaction, and detection. We will particularly focus on diamond \textit{photonic} structures coupled to single quantum systems, and not discuss diamond plasmonics ~\cite{schietinger_plasmon-enhanced_2009,barth_controlled_2010,bulu_plasmonic_2011,choy_spontaneous_2013}, non-linear photonics in diamond resonators~\cite{hausmann_diamond_2014}, Raman-lasers~\cite{mildren_highly_2009}, optomechanical systems~\cite{ovartchaiyapong_high_2012,ovartchaiyapong_dynamic_2014,khanaliloo_high-q/v_2015,ovartchaiyapong_high_2012}, and hybrid systems with diamond nanocrystals~\cite{benson_assembly_2011}. While there have been significant advances with nanodiamonds in hybrid photonic systems, and nanodiamond fabrication and properties have advanced~\cite{trusheim_scalable_2014,knowles_observing_2014}, in this paper we will focus on diamond nano- and microstructures patterned into polycrystalline and single-crystal diamond, which have been shown to have superior optical and spin properties compared to most diamond nanocrystals.

\section{Diamond for Quantum Photonics}
\label{sec:DiamondProp}

\subsection{Material Properties}
Diamond's exceptional combination of a wide electronic bandgap, high mechanical strength, high thermal conductivity, and large hole mobility has made it an attractive material for high frequency, power, temperature, and voltage applications such as power electronics~\cite{wort_diamond_2008,balmer_chemical_2009}. Furthermore, diamond is chemically inert and biocompatible making it a promising material for biological applications, especially in nanocrystalline form~\cite{mochalin_properties_2012}. In the field of quantum optics, diamond is uniquely attractive due to its wide bandgap, $\sim 5.5$~eV, which allows it to host to more than 500 optically active defects, known as `color centers'. These crystalline defects, corresponding to some combination of displacement or substitution of the native carbon atoms within the diamond lattice, create spatially localized, energetically separated ground and excited states within the electronic band gap of the bulk crystal. The high Debye temperature of diamond ($\Theta_0 \sim 2219~$K~\cite{desnoyehs_heat_1958}) leads to a relatively low phonon population at room temperature, which allows these defect-related electronic states to persist for long times without suffering from phonon-induced relaxation. Finally, diamond crystals are relatively free of nuclear spins, with a natural composition of $^{12}$C of $\sim 99\%$ that can be increased to $\sim 99.99\%$ in isotopically-enhanced chemical vapor deposition (CVD) growth~\cite{balasubramanian_ultralong_2009,teraji_chemical_2012}. This lack of background spins leads to low magnetic field fluctuations, in principal facilitating long coherence times for the few electronic or nuclear spins present. These properties make diamond an ideal host material for single quantum defects and photonic elements.

\begin{figure}[htbp]
\centering
\fbox{\includegraphics[width=\linewidth]{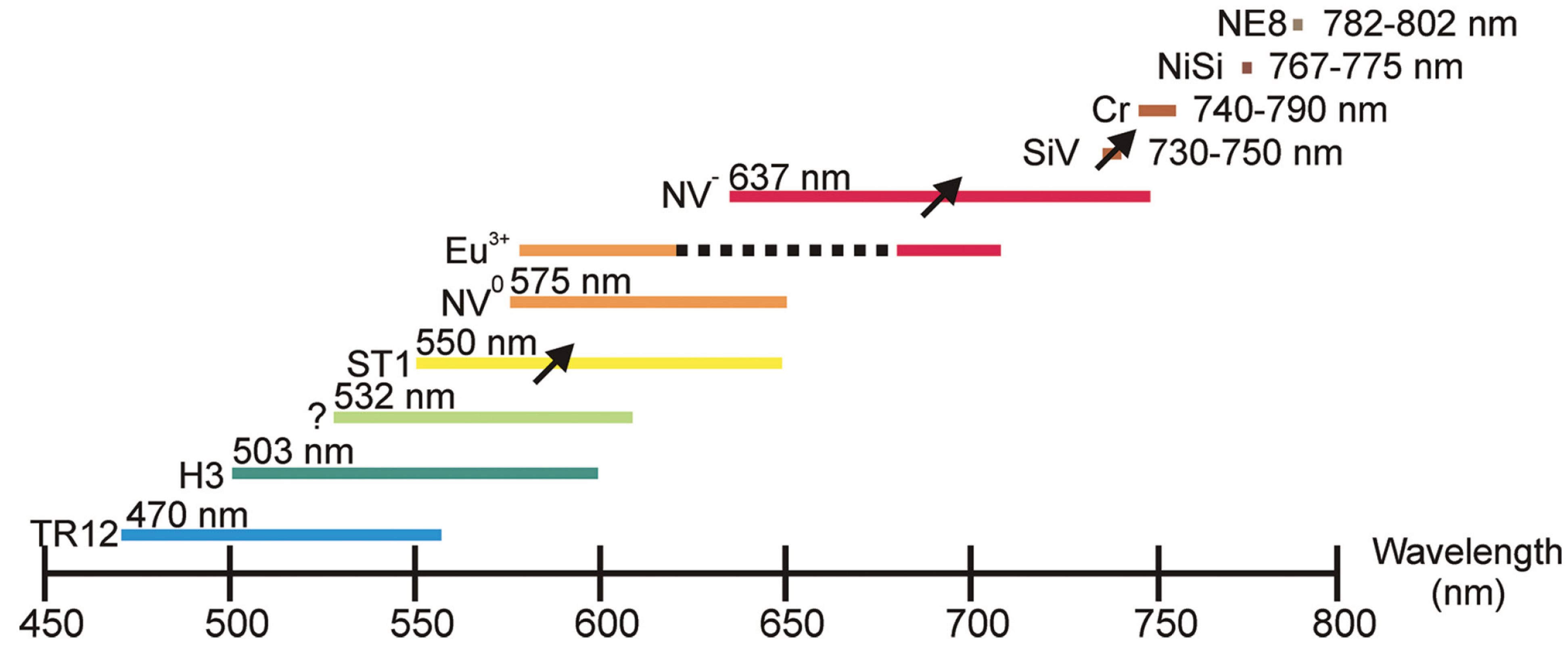}}
\caption{Overview of studied color centers. 
For centers with emission wavelengths shorter than 730 nm, the length of the colored line represents the approximate width of the emission spectrum of the color center including phonon sidebands. The wavelength given for each center
denotes the zero-phonon-line (ZPL) wavelength. The black arrows indicate that the centers spin has been addressed.
Reprinted with permission from Reference~\cite{aharonovich_diamond_2014}. \textcopyright 2014 WILEY-VCH.}
\label{fig:ColorCenters1}
\end{figure}

\subsection{Defect Centers in Diamond}
\label{sec:Defect}

Of the over 500 optically active defect centers in diamond~\cite{zaitsev_optical_2001}, more than 10 have been demonstrated to exist as single quantum emitters~\cite{aharonovich_diamond_2014}. As shown in Fig.~\ref{fig:ColorCenters1}, these centers span a wide range of single-photon emission wavelengths across the visible spectrum ranging from the blue into the near-infrared. 

Of these single-photon sources, three have been observed to exhibit optically detected magnetic resonance (ODMR), in which changes in photon emission intensity are observed while driving the spin on and off resonance with a tunable microwave field, first demonstrated for the negatively charged nitrogen vacancy center (NV$^{-}$)~\cite{gruber_scanning_1997}. This is a convenient mechanism for direct spin state readout via photoluminescence. Spin-state-dependent optical transitions, in general, enable fast initialization, manipulation and measurement of spin states using laser excitation~\cite{gao_coherent_2015}. Many quantum information processing (QIP) schemes use a link between stationary solid-state quantum memory bits and flying photonic qubits as a basic resource~\cite{kimble_quantum_2008}, making diamond spin systems an attractive candidate for QIP~\cite{awschalom_quantum_2013,wrachtrup_processing_2006}. 

The most prominent among the ODMR-active diamond color centers is the NV$^{-}$, which exhibits stable room temperature single photon emission and particularly long electron and nuclear spin coherence times compared to other solid state defect centers~\cite{doherty_nitrogen-vacancy_2013}. In the field of quantum optics, the NV$^{-}$ has notably been applied in experiments demonstrating spin-photon entanglement~\cite{togan_quantum_2010}, distant spin entanglement~\cite{bernien_heralded_2013}, quantum teleportation~\cite{pfaff_unconditional_2014}, and finally the first loophole free demonstration of Bell's inequality~\cite{hensen_loophole-free_2015}. Recently, entangled absorption was demonstrated mediated by an inherent spin-orbit entanglement in a single NV$^{-}$~\cite{kosaka_entangled_2015}. Also the coherent transfer of a photon to a single solid-state nuclear spin qubit with an average fidelity of 98\% and storage times over 10 seconds demonstrated~\cite{yang_high_2015}. In addition to the NV$^{-}$, the negatively charged silicon vacancy defect center (SiV$^{-}$) has  recently gained attention as an optically accessible single spin system. Notably, the SiV$^{-}$ in pure, strain free crystals possesses optical transitions that are naturally nearly Fourier-transform-limited and insensitive to environmental noise. This has enabled the demonstration of high-fidelity Hong-Ou-Mandel interference between photons emitted by two SiV$^{-}$ centers~\cite{sipahigil_indistinguishable_2014}. Unfortunately, the spin coherence times are currently limited by phonon interactions to several 10s of nanoseconds, which is many orders of magnitude lower than that of the NV$^{-}$.

\section{Fabrication of Diamond Photonic Devices}
\label{sec:DiamondPatterning}
The realization of diamond-based photonic devices requires that optical design parameters are accurately transferred into diamond by either an additive~\cite{furuyama_improvement_2015} or subtractive approach. We will focus on the subtractive approach, as it is more common for well designed photonic structures. Various fabrication methods have been used to demonstrate photonic devices in diamond~\cite{ando_smooth_2002,hausmann_fabrication_2010,
riedrich-moller_one-_2012,aharonovich_bottom-up_2013, sovyk_fabrication_2014,furuyama_improvement_2015,martin_subtractive_2014,tao_facile_2013}. Focused Ion Beam (FIB) milling defines and transfers the photonic pattern directly into diamond~\cite{bayn_processing_2011, lesik_maskless_2013}. Direct electron beam lithography (EBL) writing defines a pattern with nanometer precision in an electron beam resist layer and is usually combined with a subsequent dry etching step to transfer the photonic pattern into the diamond. Transferrable silicon mask lithography~\cite{li_nanofabrication_2015} exploits the relatively mature fabrication process on silicon-on-insulator (SOI) samples, and provides high etch selectivity for the subsequent oxygen etching step. Due to the lack of commercially available diamond thin films of optical thickness, 3-dimensional (3D) monolithic patterning techniques have also been developed such as angular FIB and RIE etching~\cite{burek_free-standing_2012, burek_high_2014} and isotropic etching~\cite{khanaliloo_high-q/v_2015,khanaliloo_single-crystal_2015} techniques.

\subsection{Synthetic Creation of Diamond}
All successful photonic devices must begin with high-purity single crystal diamond. While initial defect studies were done on natural diamond, current nano-photonic structures are made from synthetic diamond. Chemical vapor deposition (CVD) and high-pressure high-temperature (HPHT) growth allow for the production of low strain single-crystal diamond with controllable defect concentrations~\cite{balmer_chemical_2009,schreck_large-area_2014}. These high-quality single-crystal diamonds are limited in size to a few tens of mm$^2$, while polycrystalline films and bulk diamonds can be grown on significantly larger scales, though with less control over lattice strain and defect inclusion across the sample. Growth process parameters strongly influence the quality of the crystal and the concentration and type of lattice defects, making diamond growth still technically challenging, though high quality synthetic diamonds are available commercially. Ultra-pure diamond, categorized as type IIa for low nitrogen and boron content, is often used as a starting point for quantum optics applications because of higher purity. In fact, defects in ultra-pure diamond can be eliminated down to levels $< 1$~ppb. However, desired defects can also be introduced in the diamond growth up to high levels $> 100$~ppm, allowing for control of the native defect densities across several orders of magnitude. This approach can avoid crystalline damage associated with defect creation through implantation, and has been shown to provide long spin-coherence times~\cite{kennedy_long_2003}, high defect density or spatially-selected defect layers~\cite{ohno_engineering_2012,rogers_multiple_2014,lee_deterministic_2014} as will be discussed in Section~4c. Further insights into the synthesis of single crystal diamond by HPHT or CVD methods, either by homoepitaxial growth~\cite{teraji_homoepitaxial_2015} or heteroepitaxial deposition on large-area single crystals of a foreign material are discussed in Ref.~\cite{schreck_large-area_2014,gsell_route_2004}.

\subsection{Diamond Thin Film Fabrication}
\label{sec:sec:ThinFilm}
Many photonic systems that confine light on the order of the wavelength are based on thin film substrates which are either suspended or supported by a lower index of refraction material to achieve total internal reflection. For single mode devices operating resonantly with color centers in the visible range, such a thin film has to be on the order of 200~nm in thickness. In contrast to many other semi-conductor materials, single crystal diamond growth is challenging and has been limited mainly to diamond-on-diamond techniques, precluding the use of an underlying sacrificial layer or lower index material. 

One very promising exception is the heteroepitaxial growth via bias-enhanced nucleation~\cite{schreck_diamond_2001} 
on iridium/yttria-stabilized zirconia buffer layers on silicon (layer structure: diamond/Ir/YSZ/Si(001))~\cite{gsell_route_2004}. Due  to an unmatched degree of initial alignment and extraordinary high density of  such epitaxial diamond grains on the iridium layer, they can lose their polycrystalline character during subsequent textured growth within a few to tens of micrometers~\cite{schreck_mosaicity_2002}.
A several hundred nanometer thick suspended diamond membrane can be fabricated by first removing the silicon substrate and buffer layers and then dry-etching the polycrystalline backside of the diamond~\cite{riedrich-moller_one-_2012}. 

For the fabrication of diamond thin films from bulk diamond substrates, different methods have been developed. FIB milling has been pursued for the separation of small diamond slabs from the bulk~\cite{olivero_characterization_2006}. However, the highly physical nature of the ion bombardment causes crystal damage as evidenced by Raman spectroscopy, photoluminescence~\cite{olivero_characterization_2006}, and transmission electron microscopy~\cite{li_reactive_2013}. Reactive Ion Etching (RIE) of slabs was seen to cause less crystal damage~\cite{li_reactive_2013} in the final product, allowing for spin coherence times approaching 100\,$\upmu$s~\cite{hodges_long-lived_2012}. However, this RIE method only allowed for the production of small ( $\sim10\times10\,\upmu$m$^2$) membranes, which limits the ability to post-process and fabricate more complex photonic structures.

There has also been work towards the separation of a diamond film from the bulk via the controlled creation of a damage layer. MeV ions accelerated at the diamond crystal will stop at an energy-dependent depth. The damage caused by collisions with the lattice will create a well-localized graphite layer that can be removed via a wet-etch step~\cite{wang_fabrication_2007,fairchild_fabrication_2008}. Crystal damage is inevitably induced in the removed membrane. However, this can be mitigated with a strategic etch of the damaged side~\cite{magyar_fabrication_2011}, and subsequent diamond overgrowth~\cite{lee_fabrication_2013,aharonovich_homoepitaxial_2012,gaathon_planar_2013}, allowing for high quality spin properties of defect centers~\cite{aharonovich_homoepitaxial_2012}. 

\subsection{Focused Ion Beam Milling}
FIB milling of diamond, in which carbon atoms are mechanically removed from the lattice with accelerated Ga$^+$ ions, or O$_2$ ions is a masks process which can be used for fabrication of diamond photonic devices~\cite{bayn_triangular_2011,riedrich-moller_one-_2012,lesik_maskless_2013}. The spatial resolution is mainly limited by the ion beam width.  This gives several advantages for diamond patterning: (i) a mask is not required, eliminating the need for special handling or resist spinning, and (ii) optical isolation from the bulk is achieved simply by tilting the stage to etch at an angle relative to the beam and to undercut the structure~\cite{bayn_triangular_2011}. FIB milling has been used to demonstrate nanobeam cavities~\cite{bayn_triangular_2011} and free-standing, undercut bridge structures~\cite{Martin_Maskless_2015} in bulk diamond, and two-dimensional (2D) photonic cavities~\cite{riedrich-moller_one-_2012} in a single crystal diamond layer on a buffered Si substrate. However, this technique is limited by the relatively long milling time and inclined side walls~\cite{riedrich-moller_one-_2012} leading to limited cavity quality-factors, and the residual damage to the diamond material, which results in reduced color center properties as well as additional optical and spin background. The material damage from ion milling can be partially removed either by acid treatment and oxidation step~\cite{riedrich-moller_one-_2012} or using electron beam induced local etching~\cite{Martin_Maskless_2015}. This minimizes the optical losses and fluorescence background from the ion contamination.

\begin{figure}[htbp]
\centering
\fbox{\includegraphics[width=\linewidth]{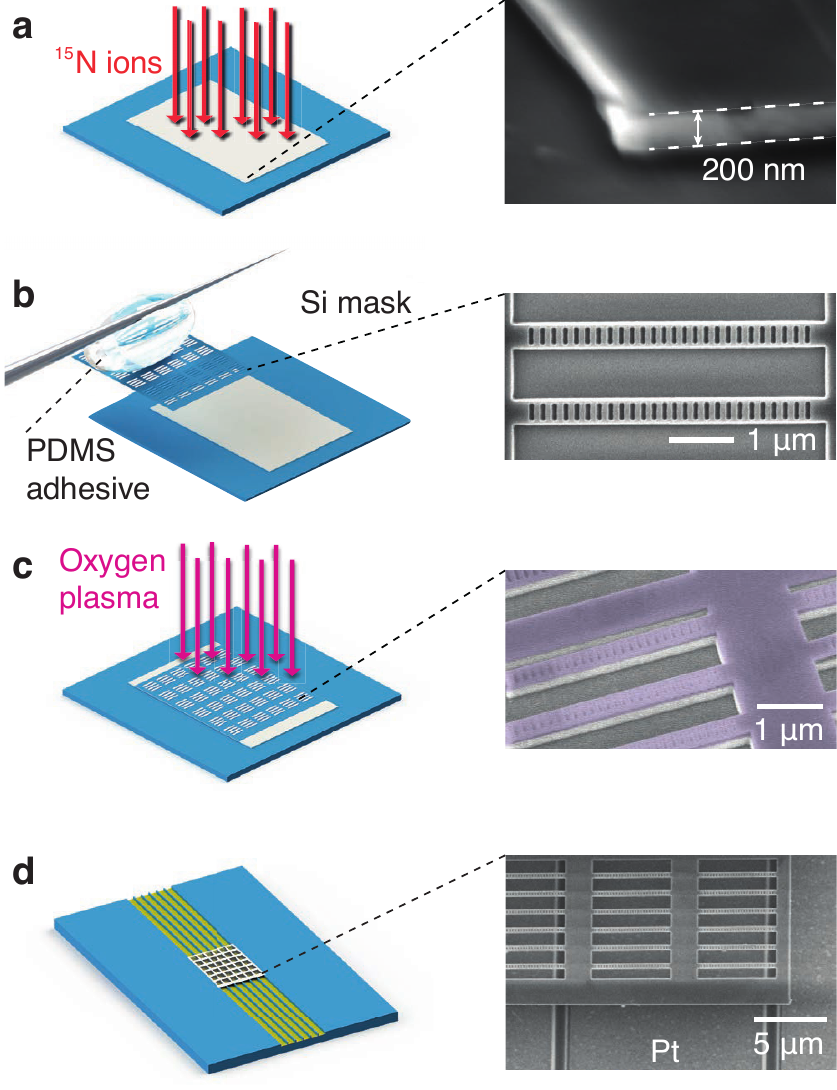}}
\caption{Fabrication procedure and cavity characterization outcome.
a) NV$^{-}$s were created $\sim100$~nm below the surface of the diamond
membranes by implantation of $^{15}$N atoms and subsequent annealing at
850$^{\circ}$C. The magnified scanning electron micrograph (SEM) shows a
$\sim200$~nm membrane. b) Silicon (Si) masks were patterned on silicon-on-insulator,
released and transferred onto diamond membranes. The SEM
shows a patterned Si mask before transfer. The scale bar represents 1~mm.
c) Oxygen reactive ion etching was used to pattern diamond membranes.
The false-color SEM shows the Si mask (purple) on diamond after oxygen
etching. The scale bar represents 1~mm. d) Patterned diamond membrane
on microwave striplines for optical and spin characterization. The SEM
shows diamond PhC structures alongside metallic striplines in Si channels.
The scale bar represents 5~mm. Reprinted with permission from Reference~\cite{li_coherent_2015}. \textcopyright 2015 Nature Publishing Group.}
\label{fig:SiPatterning}
\end{figure}

\subsection{Direct Electron Beam Lithography}
EBL is widely used for defining patterns with nanometer feature size, and is applied in combination with an etching method, most often RIE or inductively coupled plasma (ICP) RIE. EBL typically requires a conductive substrate that is several millimeters in size and an E-beam resist having sufficient etch selectivity with respect to the substrate for the subsequent pattern transfer. These requirements are challenging to satisfy for small, insulating diamond samples. Coating diamond with a conductive layer is widely used to minimize charging during EBL writing. Hydrogen silsesquioxane (HSQ) resist is a high resolution E-beam resist that can be used to pattern diamond~\cite{choy_enhanced_2011}. Its modest intrinsic selectivity to standard diamond RIE etch recipes can be enhanced by post-development electron curing~\cite{yang_enhancing_2006}. The etch selectivity can be further enhanced with other mask layers patterned via lift-off, or an initial short dry or wet etch step. Such recipes have been used to demonstrate diamond nanowires~\cite{hausmann_fabrication_2010}, suspended waveguide and nanobeam cavities~\cite{burek_free-standing_2012, burek_high_2014, lee_deterministic_2014}, diamond plasmonic apertures~\cite{choy_enhanced_2011}, and gratings~\cite{choy_spontaneous_2013}. 

\subsection{Transferrable Silicon Mask Lithography}
To avoid spin-coating small diamond samples and exposing them to electron, ion, or UV radiation, a novel nanofabrication technique~\cite{li_nanofabrication_2015} was developed as illustrated in Fig.~\ref{fig:SiPatterning}. Instead of defining a mask directly on the diamond substrate, a frameless single-crystal silicon membrane mask is pre-patterned from (SOI) samples~\cite{lipson_silicon_2007,liu_mid-infrared_2010,peroz_single_2012} and placed onto the diamond substrate silicon using membrane-transfer techniques~\cite{li_nanofabrication_2015}. This enables pattern transfer with feature sizes down to 10 nm, etch selectivity of over 38 for the subsequent oxygen RIE, and automatic positioning of ion implantation apertures with respect to the photonic structure, as will be discussed in Sec.~\ref{artificialCreation} with alignment accuracy guaranteed by the EBL writing. For patterning of the silicon masks, high resolution EBL is applied, in combination with well developed RIE. 

By applying one of two complimentary transfer techniques, one can place both small and large masks on diamond substrates with sizes up to 200~x~200~$\upmu$m, and 1~x~1~mm, as required by the sample size to be patterned. For small masks, a pick-and-place method is applied based on nanomanipulation of a micro-polydimethylsiloxane (PDMS) adhesive attached to a tungsten probe tip. For large masks, a stamping approach is used with a transparent polytetrafluoroethylene (PTFE) sheet~\cite{li_nanofabrication_2015}. After the pattern is transferred to diamond by oxygen etching, the Si membrane masks are mechanically removed, avoiding solvent-based mask removal procedures on diamond substrates.

The Si mask patterning was applied for various photonic nanostructures in both bulk samples~\cite{bayn_fabrication_2014} and diamond thin films with sizes down to hundreds of square microns~\cite{li_coherent_2015}. To pattern diamond membranes with $200-300$~nm thickness, the diamond membranes are adhered to a Si substrate, the patterned Si mask is put on the diamond membrane using the pick-and-place technique, photonic patterns are transferred into the diamond membrane by oxygen RIE, the Si mask is mechanically removed, and finally isotropic SF$_6$ plasma is applied to undercut the etched diamond photonic structure for optical measurements.

\subsection{Angular Etching and Isotropic Etching Techniques}
\label{sec:AngEtching}
As discussed above, the fabrication of large uniform thin film diamond samples has not yet been developed. This limits the fabrication of diamond devices that guide and capture light when 2D or 3D confinement is required. To achieve the needed optical isolation for photonic structures in bulk diamond, and to circumvent the need for large thin film diamond samples for patterning, alternative fabrication approaches have been demonstrated that enable the monolithic patterning of waveguide and cavity structures into bulk diamond. One design concept is based on suspended devices with triangular cross-section. Such designs can be realized by angled etching, either by rotating the sample~\cite{bayn_triangular_2011} and milling by FIB, or by guiding the trajectory of ions with a Faraday cage~\cite{burek_free-standing_2012} in an RIE process. Angular etching was used to demonstrate race-track patterns with ultra high quality factor (Q factor)~\cite{burek_free-standing_2012} and one-dimensional (1D) nanobeam cavities\cite{burek_high_2014,bayn_fabrication_2014}. 

Another technique to produce free-standing structures is a quasi-isotropic oxygen undercut. This technique is based on the combination of standard vertical RIE and zero forward bias oxygen plasma etching at an elevated sample temperature. This zero bias etching takes advantage of the low directionality of the oxygen ions and the thermally activated diamond surface leading to a quasi-isotropic chemical etching effect. This technique has enabled high-Q cavities in a nanofabricated photonic disk~\cite{khanaliloo_high-q/v_2015}, and high mechanical quality factor waveguides~\cite{khanaliloo_single-crystal_2015}.

\section{Synthetic Creation of Defect centers}
\label{artificialCreation}
 
Color centers can be found in natural or as-grown synthetic diamond. However, for high quality samples with very low defect concentrations (e.g. N\,<\,1ppm) the concentration of color centers is too low for many of the intended applications~\cite{khan_colour-causing_2013}, and the distribution is random. Controlled creation of defect centers is important for the fabrication of photonic constituents in a scalable way and for the extension beyond present proof-of-principle implementations. One can differentiate between methods that rely on (i) 'activation' of incorporated defects in as-grown diamond~\cite{martin_generation_1999}, (ii) controlled incorporation of defects during growth~\cite{ohno_engineering_2012}, and (iii) controlled implantation of defects after diamond growth~\cite{yang_single_2003}. Combinations of these methods have also been demonstrated. For example, 'activation' methods can be combined with the controlled or targeted 'placement' methods~\cite{mclellan_deterministic_2015}. A very powerful tool is the spatially deterministic creation via focused ion beam or masked implantation~\cite{bayn_generation_2015}. These methods enable high yield creation of nanostructures with incorporated defect centers.

The synthetic creation of defect centers depends on a wide range of parameters; annealing temperature, vacancy density, and local charge environment have all been shown to affect NV creation. While many works address NV formation as a function of these parameters, the detailed mechanism of color center creation is still not definitively understood. Until recently, it was commonly proposed that diffusing vacancies are trapped by substitutional atoms (e.g., nitrogen) to create a color center (e.g., the NV)~\cite{davies_charge_1977,acosta_diamonds_2009,pezzagna_creation_2010,orwa_engineering_2011}. Therefore, the established recipes for creating defect centers rely on annealing above 600$^{\circ}$C, at which temperature vacancies become mobile~\cite{davies_optical_1976,collins_annealing_2009}. This mechanism has been questioned by advanced density functional theory (DFT) calculations that were applied to determine the formation and excitation energies, the charge transition levels, and the diffusion activation energies for nitrogen- and vacancy-related defects in diamond~\cite{deak_formation_2014}. These calculations concluded that irradiation of diamond is more likely to directly create NV defects, and not isolated vacancies. Direct NV creation has been shown without thermal annealing by irradiation of diamond that has been implanted with nitrogen ions with low-energy electrons ($2-30$~keV)~\cite{schwartz_effects_2012} and beams of swift heavy ions ($\sim$1~GeV, $\sim$4~MeV/u)~\cite{schwartz_local_2014}. However, this model of direct NV creation is contradicted by other works. For example, experimental results still show evidence of high vacancy mobility and indicate formation of NVs after implantation during annealing~\cite{santori_vertical_2009,ohno_three-dimensional_2014}. Further fundamental investigation of defect center creation is required to understand this process in full detail.

\subsection{Annealing Temperature and Optical Properties}
Annealing temperatures are a crucial tool to control the concentration of different types of lattice and crystal defects. While it is in principle sufficient to anneal samples just above above 600$^{\circ}$C, temperatures around 850$^{\circ}$C were chosen for most demonstrations over the past years. Recently, temperatures up to 1200$^{\circ}$C are being applied to reduce strain and lattice defects, leading to increased spin-coherence times of NV$^{-}$s~\cite{naydenov_increasing_2010}. For the SiV$^{-}$ this leads to a narrowing of the inhomogeneous distribution from $3-4$~nm (after 800$^{\circ}$C anneal) to 0.03~nm (15~GHz, after 1100$^{\circ}$C anneal), and results in nearly lifetime-limited optical linewidths~\cite{evans_coherent_2015}. Furthermore, above 1100$^{\circ}$C the concentration of di-vacancies is reduced as their bonds are broken. Di-vacancies are suspected to influence the photostability of NV$^-$ centers~\cite{deak_formation_2014} and spectral diffusion properties of the NV$^{-}$ ZPL. For N implantation doses of $10^9$/cm$^2$, energies of 85~keV and annealing up to 1200$^{\circ}$C, stable optical transitions of the NV$^{-}$ ZPL with linewidth down to 27~MHz have been demonstrated~\cite{chu_coherent_2014}, which is close to the lifetime-limited emission linewidth of about $\sim$13~MHz~\cite{tamarat_stark_2006}. Charge-state stability has been shown to be directly effected by surface termination, with fluorination leading to a higher concentration of stable NV$^-$ centers~\cite{cui_effect_2013}. It is commonly assumed that surface treatment also affects the NV$^{-}$ ZPL stability~\cite{chu_coherent_2014}. However this has not been systematically studied in the literature at the time of this review.

\subsection{Activation of Incorporated Ions in As-Grown Diamond}
Defect centers can be formed by the creation of additional vacancies in doped diamond by irradiation with energetic neutrons, electrons, or ions ~\cite{mainwood_nitrogen_1994,martin_generation_1999,waldermann_creating_2007} in combination with a subsequent annealing step above 600$^{\circ}$C. In this process, ions already present in the diamond lattice can be combined with the newly created vacancies. 

Early work used electron and Ga$^+$ beams to irradiate N-rich type-Ib diamond to create vacancies and indirectly defect centers, in particular NV centers, from already incorporated N ions. For an unpatterned diamond surface, a spatial lateral resolution below 180~nm was achieved~\cite{martin_generation_1999}. Controlling the creation depth relative to the surface is challenging, as lattice defects are created along the path of the particle in the lattice, and the scattering cross-section varies for every species. Scanning focused He-ion irradiation and subsequent annealing was also applied for the creation of NV centers~\cite{huang_diamond_2013,mccloskey_helium_2014}. While these works achieve spatially localized NV creation, large areas, in particular entire samples, can also be irradiated to create large ensembles. Such large area irradiation was, for example, used to create a millimeter-scale diamond sample with about 16~ppm  (corresponding to $2.8 * 10^{18}~cm^{-3}$) NVs~\cite{acosta_diamonds_2009}. Such samples with large ensembles of spins enable magnetic-field measurements with sensitivities down to $< 0.5~\text{nT} ~{\text{Hz}^{-1/2}}$ in the low-frequency regime around 1~Hz ~\cite{le_sage_efficient_2012,clevenson_broadband_2015}. For effective sensor volumes of $8.5 \times 10{-4}$~mm$^3$ and ensembles of $\textrm{N}\sim10^{11}$ NV$^{-}$s, photon-shot-noise-limited magnetic-field sensitivity was demonstrated with a sensitivity of 0.9~pT$/\sqrt{Hz}$ for ac signals of $f=20$~kHz.

\subsection{Controlled Incorporation of Ions during Growth: Delta Doping}
\label{Delta}
An alternative way of controlling the depth of defect centers relative to the diamond surface is delta-doping. This has been experimentally demonstrated for the NV$^{-}$~\cite{ohno_engineering_2012} and the SiV$^{-}$~\cite{rogers_multiple_2014}. For the NV, a nanometer-thick nitrogen-doped layer is created by the controlled introduction of N$_2$ gas during plasma enhanced chemical vapor deposition (PECVD) diamond growth. Similarly, SiV$^{-}$s are created by controlling the Si concentration during the growth, giving control of the concentration over two orders of magnitude. Subsequent electron irradiation and annealing leads to formation of NV centers in a thin layer. This final NV creation process causes less crystal damage than direct ion implantation methods~\cite{naydenov_increasing_2010}, and is therefore advantageous for both long spin-coherence times and stable and narrow spectral linewidth. Delta-doped diamond thin-films have been applied to couple ensembles of NV$^{-}$s to a nanobeam photonic crystal cavity, demonstrating that this technique could be interesting for single-NV cavity coupled systems~\cite{lee_deterministic_2014}. However, in this work confinement was demonstrated in 1D only, but could be combined with FIB or ebeam 'activation' to achieve 3D confinement. For high-resolution sensing in fluids, delta-doping enabled engineered diamond probes with diameter and height ranging from 100\,nm to 700\,nm and 500\,nm to 2\,$\upmu$m, respectively~\cite{andrich_engineered_2014}. 

Besides incorporating atomic defects into the lattice, the concept of delta doping can also be applied to 
engineer specific nuclear spin environments, e.g., nanometer-thick layers of $^{13}$C in ultra-pure natural abundance $^{13}$C diamond by switching between purified $^{12}$CH$_4$ and $^{13}$CH$_4$ (99.99\%) source gases~\cite{ohno_engineering_2012} during diamond PECVD growth. This is promising for the creation of a controlled number or distribution of $^{13}$C nuclear spin memories, which could be used for spin-spin entanglement, quantum error correction protocols or for quantum simulators~\cite{gaebel_room-temperature_2006,balasubramanian_ultralong_2009,cai_large-scale_2013,waldherr_quantum_2014,taminiau_universal_2014}.

\begin{figure}[htbp]
\centering
\fbox{\includegraphics[width=0.98\linewidth]{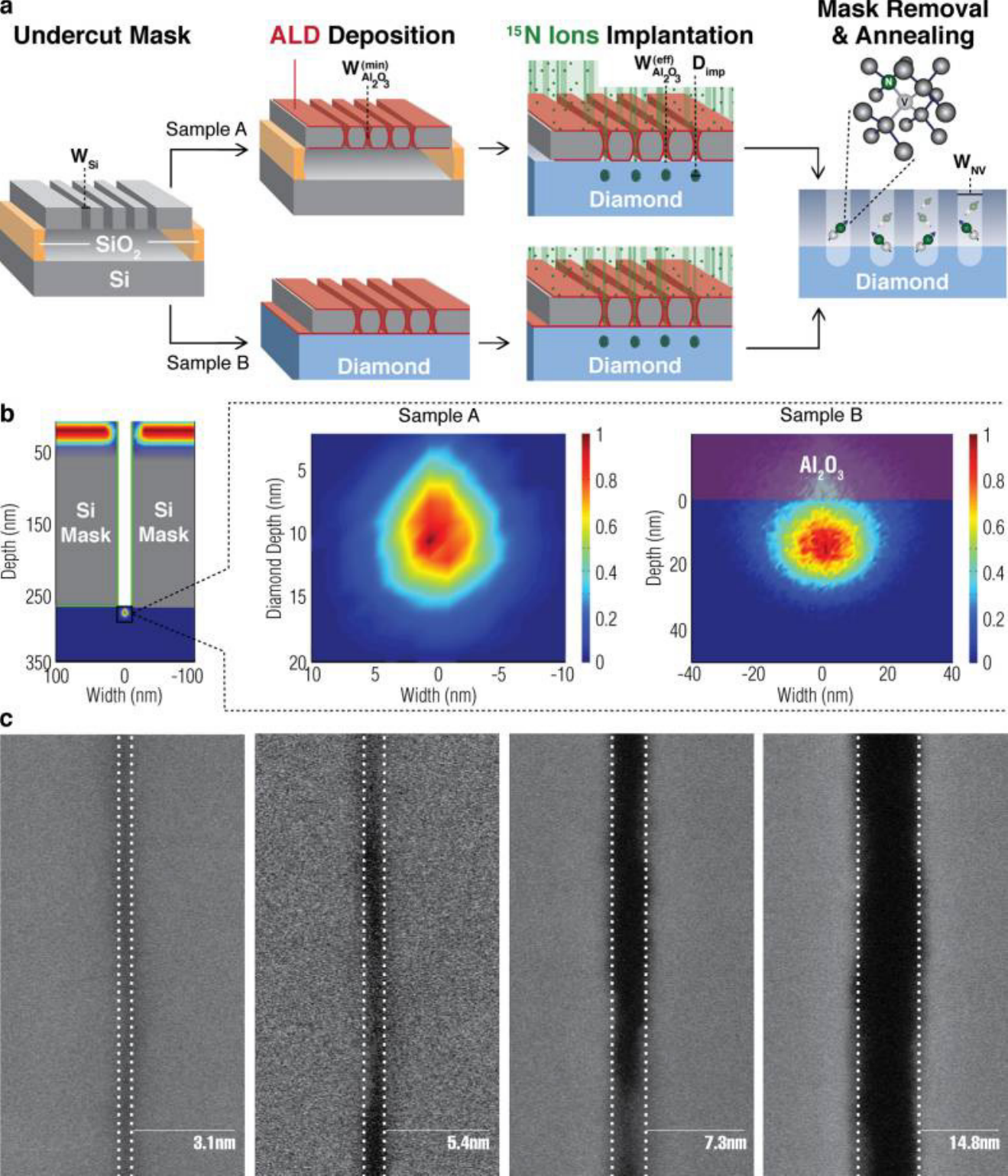}}
\caption{Implantation through silicon hard mask. a) Fabrication scheme from silicon mask undercut to implantation for Samples A and B. b) SRIM simulation of implanted nitrogen. Vertical cross section of the mask and implantation region (left); close-up of nitrogen post-implantation ion on Sample A (center, 6~keV) and Sample B (right, 20~keV) with lateral straggle of 3.1 and 11~nm, respectively. Color bar: normalized nitrogen density. c) Top-down SEM images of mask following ALD, showing a minimum mask width of 3.1~nm.
Reprinted with permission from Reference~\cite{bayn_generation_2015}. \textcopyright 2015 ACS Publications.}
\label{fig:TargetCreation}
\end{figure}

\subsection{Implantation of Ions}
The most common method for the creation of color centers is direct ion implantation of a color center constituent -- for example, N, Si, Cr, or other ions~\cite{yang_single_2003,meijer_generation_2005,rabeau_implantation_2006,pezzagna_creation_2011} -- into the diamond crystal. Subsequent annealing creates defect centers. This method enables control of the center depth with respect to the surface, as can be determined via SRIM simulations~\cite{ziegler_stopping_1985,ziegler_interactions_????}. The first demonstration of this method used N ions with $\sim$2~MeV energy, corresponding to an implantation depth of about 1.15~$\upmu$m~\cite{meijer_generation_2005}.
Shallow NV$^{-}$s were created with 7~keV ion energy, corresponding to an implantation depth of about 10~nm, and the typical $^{15}$N 2-fold hyperfine splitting was demonstrated, in contrast to a 3-fold hyperfine splitting for natural $^{14}$N defects~\cite{rabeau_implantation_2006}. Single photon emission was demonstrated from single SiV$^{-}$ centers created via ion implantation of $^{29}$Si ions~\cite{wang_single_2006}. To further understand the creation process of color centers via ion implantation, the creation efficiency of NVs as a function of ion energy was experimentally determined~\cite{pezzagna_creation_2010} and compared to theoretical models, leading to 25\% NV creation yield per implanted nitrogen ion~\cite{antonov_statistical_2014}. All these examples have not demonstrated control of the lateral position of created color centers. 

\subsection{Targeted Creation of Defect Centers}
The relative alignment of color centers to each other is important for the deterministic arrangement in an array, in particular if these centers are used as a grid of sensors or as a network of entangled spins. Depending on the application, lattice constants as low as tens of nanometers are required with precise positioning at each lattice site. One way to achieve this goal is the targeted implantation through 30~nm apertures in the tip of an atomic force microscope (AFM)~\cite{meijer_towards_2008}. This AFM method was combined with stimulated emission depletion microscopy~\cite{han_three-dimensional_2009} to demonstrate nanometer-scale mapping of randomly distributed NV$^{-}$s within a less than 100~nm diameter spot~\cite{pezzagna_nanoscale_2010}.

A different method for the precise relative alignment of color centers is the implantation through large-scale lithographically defined apertures, for example, EBL written apertures in beam resist ~\cite{toyli_chip-scale_2010} or EBL patterned Si-masks. In the latter experiment, 
ensembles of individually resolvable NV$^{-}$s were created with nanometer-scale apertures in ultrahigh-aspect ratio implantation masks. These masks were fabricated by narrowing down apertures via atomic-layer-deposition (ALD) of alumina, enabling a Gaussian FWHM spatial distribution of about 26.3~nm, thus, reaching the lateral implantation straggle limit~\cite{bayn_generation_2015}, see Fig.~\ref{fig:TargetCreation}.

Irradiation is not limited to as-grown diamond but can also be used to increase the creation yield of delta-doped samples. For example, $^{12}$C implantation of a delta-doped sample post-growth, creates additional lattice defects, and individual NV$^{-}$s can be localized within a volume of (180~nm)$^3$ in an unpatterned diamond at a predetermined position defined by an implantation aperture~\cite{ohno_three-dimensional_2014}. Alternatively, by combining delta-doping for vertical confinement, and electron irradiation in a transmission electron microscope (TEM) for lateral confinement, NV$^{-}$s were created in a volume of less than 4~nm~x~1~$\upmu$m$^2$~\cite{mclellan_deterministic_2015}.

A very versatile tool for the creation of color center arrays is focused ion beam implantation. It is maskless and enables the implantation of almost arbitrary patterns. Similar to the electron beam in an scanning electron microscope, a focused ion beam can be applied to control the position and concentration of
ions. This method has been used to implant N ions within a spot size of approximately 100 nm~\cite{lesik_maskless_2013}. Similarly, arrays of silicon-vacancy centers were created by low-energy focused ion beam implantation~\cite{tamura_array_2014}. These are promising methods towards the targeted coupling of single defect centers to nanostructures.

\subsection{Deterministic Coupling To Optical Cavities}
\begin{figure}[htbp]
\centering
\fbox{\includegraphics[width=\linewidth]{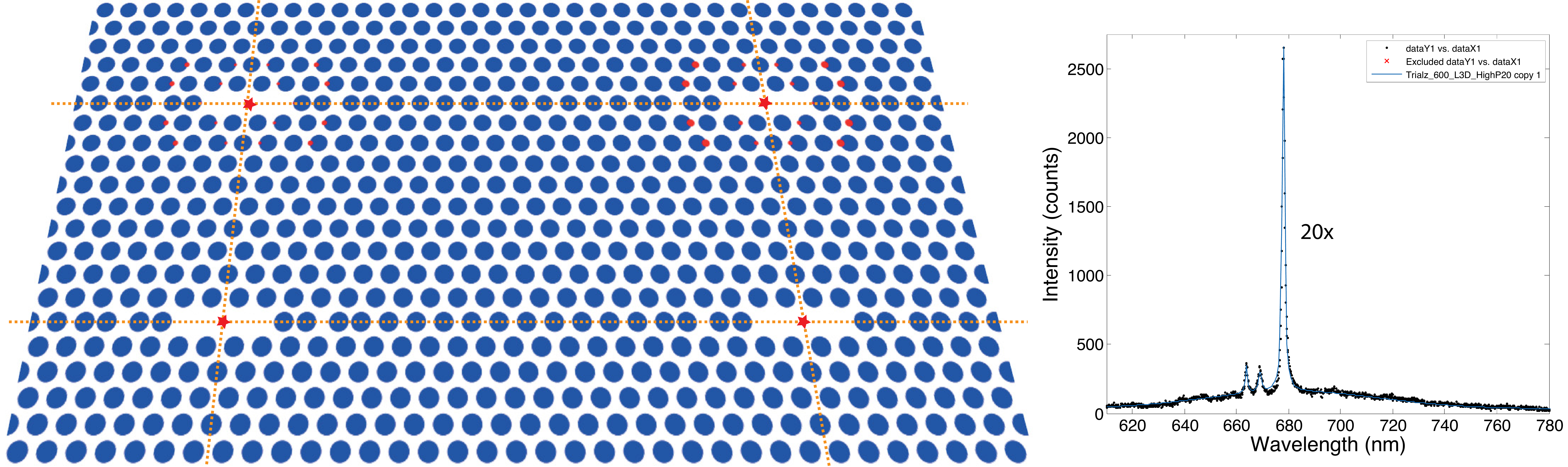}}
\caption{Targeted creation of single NV$^{-}$ defect centers in the mode maximum of L3~photonic crystal cavities. The illustration indicates an ideal NV$^{-}$-cavity system: In the center of each cavity, a single NV$^{-}$ was created. The spectrum is an example of the experimental implementation that yielded in a high number of NV$^{-}$-cavity systems. For certain fabrication runs, most of the cavities in large arrays of more then 100 cavities were coupled to a single or few NV$^{-}$s and showed intensity enhancement of the phonon sideband by a factor of 5 to 20~\cite{schroder_targeted_2014}.
Adapted reprint with permission from Reference~\cite{schroder_deterministic_2015}. \textcopyright 2015 OSA Publications.
}
\label{fig:TargetCreationL3}
\end{figure}

The deterministic coupling of a single or few color centers to a photonic nanostructure, in particular a high-Q cavity, is one of the most important prerequisites for the upscaling of QI architectures. Such deterministic coupling was recently demonstrated by fabrication of a photonic crystal cavity around a pre-characterized SiV$^{-}$ by FIB milling~\cite{riedrich-moller_deterministic_2014} enabling a resonant Purcell enhancement of the zero-phonon transition by a factor of 19, mainly limited by the positioning accuracy. To achieve a higher positioning accuracy, targeted implantation of ions into a photonic crystal cavity was realized with the AFM method discussed earlier, and Purcell enhancement of single NV$^{-}$ centers was demonstrated~\cite{riedrich-moller_nanoimplantation_2015}. A more scalable fabrication method for cavity-defect center systems is based on an implantation mask with small apertures of $30-70$~nm in diameter for targeting a large number of cavity mode maxima with a wide $^{15}$N beam~\cite{schroder_targeted_2014}. By combining the nanocavity etch mask with an implantation mask into a single physical mask, RIE etching (see Sec.~\ref{sec:DiamondPatterning}) and implantation can be carried out subsequently without the need of challenging re-alignment processes for two-mask processes. With this method, intensity enhancement of a factor up to 20 was demonstrated (Fig.~\ref{fig:TargetCreationL3})~\cite{schroder_deterministic_2015}.\\

\subsection{Dipole Orientation of Defects}
To achieve optimal light-matter coupling in cavities and other photonic systems, the defect centers must not only have precise spatial positing with respect to the optical field, but their emission dipoles must also be correctly aligned. Each color center has an individual atom-vacancy composition and geometry, and therefore a certain emission dipole orientation. For instance, the NV center has four possible orientations with respect to the crystal lattice. Naturally occurring NV populations have a random distribution of orientations. In the last few years, research has been done to control the orientation of NV centers during CVD growth and thus increase the device yield. Initial studies showed that diamond grown with homoepitaxial CVD growth on [110] ~\cite{edmonds_production_2012} and [100]~\cite{pham_enhanced_2012} oriented substrates mainly supports two NV orientations when the growth parameters are controlled precisely. Further work showed that for NV centers created during CVD of diamond on [111] surfaces, microwave plasma-assisted CVD yields 94\% of NV centers along a single crystallographic direction~\cite{michl_perfect_2014}. A different study showed that 97\% perfect alignment can be obtained by controlling the CVD growth parameters precisely~\cite{lesik_perfect_2014}. This research is promising, and if the exact mechanisms can be understood and these samples can be made routinely with high yield, this technology can help increase quantum sensing sensitivity as well as interaction with fabricated photonic structures.

\section{Diamond Devices I: Enhanced Light Collection}
\label{sec:DiamondPhotonic}

In this section, we will focus on micro- and nanophotonic structures to increase the collection efficiency of photons emitted by defect centers. A higher collection efficiency leads to improved entanglement rates for both emission-based and absorption-based quantum communication applications and also to higher read-out fidelities for quantum sensing applications~\cite{nemoto_photonic_2014,shields_efficient_2015}. Without any modification of the diamond surface, the high refractive index ($\sim2.4$) of diamond results in a relatively small angle ($\sim24.6^{\circ}$) of total internal reflection at diamond-air interfaces, allowing only a small part of the overall emission to exit the diamond. 
Even for very shallow NV$^{-}$s (several 10s of nm in depth) relevant for sensing applications, emission into the air is unfavorable due to the directed emission of a dipole into the higher refractive index material~\cite{lukosz_light_1977,lee_planar_2011,schroder_ultrabright_2011}. To overcome this limitation, a variety of photonic structures have been implemented at the diamond-air interface. A selection of devices and methods is discussed in more detail in this section.

\subsection{Waveguiding Structures}
\label{subsec:Nanowire}
One approach for overcoming limited collection efficiencies at diamond-air interfaces are cylinder- or cone-like structures etched into diamond. Depending on their shape and aspect ratio, they are referred to as diamond microcylinders~\cite{ando_smooth_2002}, nanowires~\cite{babinec_diamond_2010,hausmann_fabrication_2010}, nanopillars~\cite{neu_photonic_2014}, nanobeams~\cite{shields_efficient_2015}, or nanowaveguides~\cite{momenzadeh_nanoengineered_2015}.

Conceptually, these structures are micro- or nanometer size single-mode waveguides: the defect couples directly to a single waveguide (WG) mode, while emission into other modes is suppressed. This enables efficient coupling to that specific WG mode. For a relatively narrow emission line of a few nanometers, e.g. the ZPL of an NV$^{-}$ or SiV$^{-}$, the coupling efficiency can be up to 86\%~\cite{mouradian_scalable_2015}. From the WG mode the light is then either launched into free-space, bulk diamond, or another guiding photonic structure, enabling high overall collection efficiencies. Such structures can be used as standalone devices as discussed in this section or can be integrated in hybrid photonic circuits and fiber architectures (Sec.~\ref{sec:Hybrid}).

The first demonstration of microcylinders with the high aspect ratio of 8 (25 for exceptional cases) did not yet consider  photonic applications but demonstrated smooth and high-rate reactive ion etching of diamond~\cite{ando_smooth_2002}. Numerical modelling was later used to study the coupling of an NV$^{-}$ to the optical modes of a nanowire, and to determine the optimal nanowire parameters for large photon collection efficiency. For nanowires with diameters of 180~nm to 230~nm and for s-polarized dipoles with nanometer emission linewidth, more than 80\% of emitted photons can couple to the nanowire mode~\cite{hausmann_fabrication_2010}. Such nanowires were realized in the same work in both bulk single crystal and polycrystalline diamond and were applied to demonstrate high photon collection efficiencies of an NV$^{-}$ with a detected photon flux of about $168\pm 37$~kcts/s, ten times greater than for bulk diamond while using ten times less laser excitation power ($\sim$58~$\upmu$W under the same excitation conditions (objective lens with a numerical aperture of 0.95)~\cite{babinec_diamond_2010}. An improvement to about $304$~kcts/s photon flux in saturation was achieved with NV$^{-}$s located about $\sim1~\upmu$m away from the nanowire end by combining ion implantation and top-down diamond nanofabrication~\cite{hausmann_single-color_2011}.

These experiments were realized on [100]-oriented diamond, where the NV$^{-}$ dipole is inclined to the nanowire axis, limiting the NV$^{-}$ dipole coupling to the nanowire mode, hence limiting the overall photon collection efficiency. This limitation was overcome by fabricating such nano-structures on [111]-oriented diamond for which the electric dipole can be in-plane, increasing saturated fluorescence count rates to over $10^6$ counts per second~\cite{neu_photonic_2014}. Spin coherence measurements of NV$^{-}$s in the sample before and after fabrication demonstrated the quality of their nano-fabrication procedure, with average spin coherence times remaining unaffected at $\sim200\,\upmu$s. A further study in controlling the shape of nano pillars  and its corresponding guiding properties was realized with EBL and inductively coupled plasma RIE with a two resin technology and the usage of a titanium metal mask~\cite{widmann_fabrication_2015}.

By integrating a nanopillar into a diamond AFM cantilever and additionally positioning a single NV$^{-}$ at the tip of the nanopillar close to the diamond surface, a robust scanning sensor for nanoscale imaging was realized, demonstrating imaging of magnetic domains with widths of 25 nm, and magnetic field sensitivities down to 56~nT/Hz$^{1/2}$ at a frequency of 33~kHz~\cite{maletinsky_robust_2012}. For high-resolution sensing in fluid, cylindrical diamonds particles with diameter (height) ranging from 100\,nm to 700\,nm (500\,nm to $2~\upmu$m), were fabricated with shallow-doped NV$^{-}$ centers~\cite{andrich_engineered_2014}. The defects in these nanostructures retained spin coherence times >700\,$\upmu$s, enabling an experimental DC magnetic field sensitivity of 9\,$\upmu$T$/\sqrt{\text{Hz}}$ in fluid.	

Nanobeam waveguides with a triangular cross-section of 300\,nm width and 20\,$\upmu$m length were fabricated as free-standing structures in bulk diamond with an angled reactive ion etching~\cite{burek_free-standing_2012}, and were placed on a cover slip for oil immersion spectroscopy. By adding 50-nm deep notches every 2$\upmu$m along the beam, guided light is scattered to be collected with high numerical aperture (NA=1.49) collection optics, leading to saturation photon count rates of about 0.95~Mcts/s~\cite{shields_efficient_2015} . These structures were used to demonstrate 
efficient spin readout of NV$^{-}$ centers based on conversion of the electronic spin state of the NV$^{-}$ to a charge-state distribution, followed by single-shot readout of the charge state~\cite{shields_efficient_2015}.

An asymmetric waveguide design was demonstrated for even higher collection efficiencies for monolithic bulk structures and close to the surface sensing applications~\cite{momenzadeh_nanoengineered_2015}. These pillar-shaped nanowaveguides have a top diameter of 400~nm and a bottom diameter of up to 900~nm. This variation modifies the effective refractive index along the pillar as well as the propagation constant for each mode~\cite{love_tapered_1991}, enabling saturation photon count rates of up to $\sim 1.7$~Mcts/s. The temperature dependency of the T1 relaxation time of a single shallow NV$^{-}$ electronic spin was determined with this structure.

\subsection{Solid Immersion Lenses}
In contrast to collecting the defect emission via coupling to a waveguide mode, solid immersion lenses (SILs) enable efficient outcoupling from bulk diamond by providing perpendicular angles of incidence at the diamond-air interface. In the simplest implementation, the emission pattern from a color center in diamond is not altered, but a higher fraction of light is emitted into the free-space.  Although so-called Weierstrass~\cite{yoshita_improved_2002}, and elliptical~\cite{schell_numerical_2014} designs promise higher collection efficiencies compared to the standard hemispheric shape, in diamond only hemispheric designs have been realized. One can differentiate between 
microscopic~\cite{choi_fabrication_2005} (a few to a few ten $\upmu$m in size) and macroscopic SILs~\cite{schroder_ultrabright_2011}.

Microlens arrays were first realized by fabrication of natural diamond by a combination of photoresist reflow and plasma etching with lens diameters ranging from 10 to 100~$\upmu$m~\cite{choi_fabrication_2005}. Concave and convex microlenses with diameters ranging from 10 to 100~$\upmu$m were fabricated with hot-embossing and photoresist re-flow, followed by ICP etching were applied~\cite{lee_fabrication_2006}.

Microscopic lenses offer the advantage of micro-integration and straight-forward fabrication via FIB~\cite{hadden_strongly_2010} enabling fabrication around a pre-characterized defect centers~
\cite{marseglia_nanofabricated_2011}. However, their extraction efficiencies are more sensitive to surface roughness and non-ideal shapes than macroscopic lenses. Still, a roughly 10-fold enhancement of the photon detection rate was achieved with 5~$\upmu$m SILs~\cite{hadden_strongly_2010}. Recently, by further optimizing FIB fabrication and alignment parameters, position accuracies of better than 100~nm (lateral) and 500~nm (axial) were demonstrated, leading to  saturation count rates of about 1~Mcts/s for a single NV$^{-}$ center oriented perpendicular to the [111] cut diamond surface ~\cite{jamali_microscopic_2014}. The deterministic alignment relative to color centers, high collection efficiencies, and relative ease of fabrication have made SILs a valuable tool for QI protocols where efficient photon collection is of crucial importance, for example, quantum interference experiments for the heralded entanglement of distant NV$^{-}$ qubits~\cite{bernien_heralded_2013} and unconditional quantum teleportation between them~\cite{pfaff_unconditional_2014}.

SILs were also demonstrated for the SiV$^{-}$ center, enabling higher photon collection efficiencies for fundamental investigations of the electronic structure of the SiV$^{-}$~\cite{hepp_electronic_2014} and for the demonstration of multiple spectrally identical SiV$^{-}$ with spectral overlap of up to 91\% and nearly transform-limited excitation linewidths~\cite{rogers_multiple_2014}.

A macroscopic 1~mm in diameter diamond SIL with surface flatness better than 10 nm (rms) was fabricated with combination of laser and mechanical processing stages, leading to a saturation count rate of 493~kcts/s from a single NV$^{-}$~\cite{siyushev_monolithic_2010}. Macroscopic SILs from other materials with high refractive index such as gallium phosphite (GaP, $\textrm{n}=3.3$) can also be applied to bulk or thin film diamond if the surfaces are smooth and flat enough to prevent airgaps of more than a few 10 nanometers. For thin film diamond, this leads to a more efficient collection due to the asymmetric refractive index profile around the color center (one side GaP, on the other side air), providing a broadband antenna mechanism for color centers. For a single NV$^{-}$, saturation count rates of 633~kcts/s were demonstrated~\cite{riedel_low-loss_2014}.

\begin{figure}[htbp]
\centering
\fbox{\includegraphics[width=\linewidth]{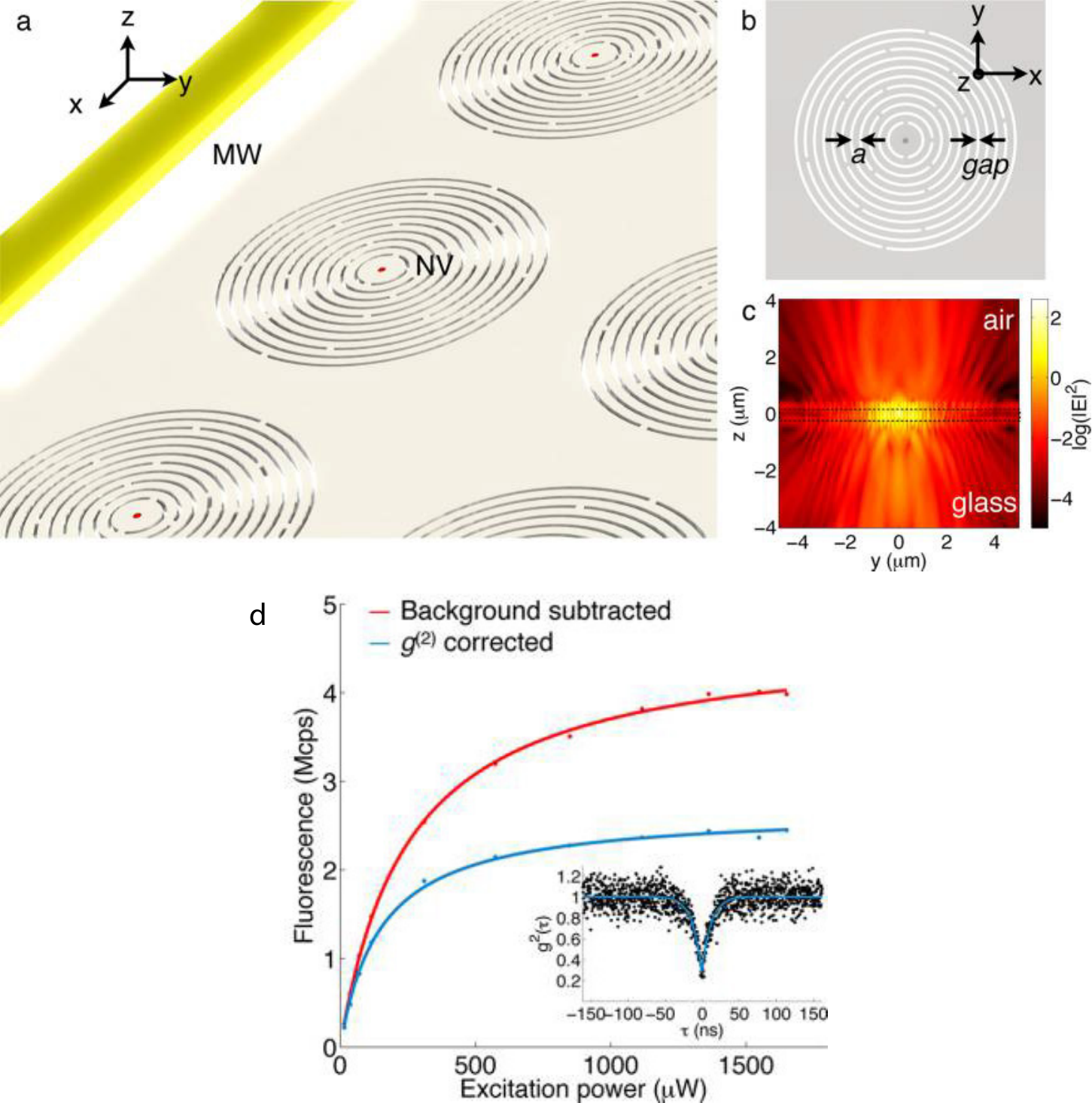}}
\caption{Bullseye grating structure for high collection efficiency of single defect center emission. a) Illustration. b) Parameters of the structure, $a$ denotes the lattice constant, and $gap$ the air spacing between circular gratings. c) Finite difference time domain (FDTD) simulation of the electric field intensity in the x = 0 plane with refractive indices corresponding diamond (bullseye), to air  (above), and glass (below). Here, $a = 330$~nm, and $gap = 99$~nm. A dipole emitter was placed in the center of the bullseye grating and was oriented along the horizontal direction. d) Saturation measurements from a single NV$^{-}$ enhanced by the 'bullseye' grating structure in diamond membranes.
Reprinted with permission from Reference~\cite{li_efficient_2015}. \textcopyright 2015 ACS Publications.}
\label{fig:bullseye1}
\end{figure}

\subsection{Circular Grating Structure}
To further increase the collection efficiency and overall single photon count rates from stand-alone photonic devices, a circular `bullseye' grating structure fabricated in a diamond membrane was placed directly on a glass coverslip~\cite{li_efficient_2015}, as indicated in Fig.~\ref{fig:bullseye1}. The periodic grating structure leads to constructive interference of the membrane-guided emission into the out-of-plane direction. Finite difference time domain (FDTD) simulations indicate (Fig.~\ref{fig:bullseye1}c) that up to 70\% of the ZPL emission of a horizontally oriented dipole emitter is guided into the glass coverslip, aided in part by the higher refractive index contrast of the diamond-air interface~\cite{schroder_ultrabright_2011}. Fabrication of these devices is carried out with the methods discussed in Section~\ref{sec:DiamondPatterning}. The bullseye gratings were analyzed in a home-built confocal microscope setup ( NA$=1.3$, Nikon Plan Fluor), and two methods are applied to determine the upper and lower bounds of the saturated single photon detection rates. As upper (lower) bound, a single photon collection rate of about 4.56~Mcps (2.70~Mcps) at saturation was determined. The saturation curves are plotted in Fig.~\ref{fig:bullseye1}b. Moreover, the high quality fabrication preserves the spin properties of the included NV$^{-}$ centers, with measured electron spin coherence times of 1.7$\pm$0.1s~\cite{li_efficient_2015}.

\section{Diamond Devices II: Optical Cavities}

Optical resonators enable control of the spectral emission properties of optical emitters and enhancing light-matter interaction of single spin systems is enabled by optical resonators. Applying the concepts of optical resonators to diamond photonics allows the tailoring of the light emission properties of defect centers, enhancing their fluorescence emission rates, and establishing efficient spin-photon interfaces, particularly important to correlate single spin states with single quantum states of light. There is also a proposal to improve the efficiency and fidelity of the ground state spin of an NV$^{-}$ spin using cavity-enhanced reflection measurements~\cite{young_cavity_2009}. A large variety of resonator types ranging from micro- to nanoscopic designs have been introduced, such as whispering gallery resonators, microscopic open cavity designs, and photonic crystal cavities. A conceptual overview of different cavity designs can be found in a recent review article~\cite{vahala_optical_2003}. The relevant cavity parameters are the cavity quality factor $Q\sim\lambda /\delta \lambda$ and the cavity mode volume $V_{\textrm{mode}}$ which directly influence the dipole-cavity interaction, e.g. the spontaneous emission rate enhancement $F_{\textrm{ZPL}}$ is proportional to $V_{\textrm{mode}}\sim(\lambda/n)^3$. For a large $F_{\textrm{ZPL}}^{\textrm{max}}$ we will focus on photonic crystal (PhC) nanocavities, as they enable small mode volumes~\cite{tiecke_nanophotonic_2014}, $V_{\textrm{mode}}\sim(\lambda/n)^3$, and large quality factors $Q$.

An emitter-cavity system can be described by the Jaynes-Cummings model in the Markov approximation~\cite{meystre_elements_2007}, and the Purcell factor quantifies the emitters spontaneous emission (SE) suppression or enhancement~\cite{purcell_spontaneous_1946}. In the strong Purcell regime, in which the emitter is coupled mainly to one optical mode, the SE can be significantly enhanced and the overall Purcell enhancement exceeds one ($F > 1$)~\cite{su_towards_2008,su_high-performance_2009}.

When the NV$^{-}$ ZPL is coupled to a cavity the SE rate is enhanced according to:
 \begin{equation}
F_{\textrm{ZPL}} = \xi F_{\textrm{ZPL}}^{\textrm{max}}\frac{1}{1+4Q^2(\lambda_{\textrm{ZPL}}/\lambda_{\textrm{cav}}-1)^2} \\
\end{equation}
\noindent where $F_{\textrm{ZPL}}^{\textrm{max}}=\frac{3}{4\pi^2}\left(\frac{\lambda_{\textrm{cav}}}{n}\right)^3\frac{Q}{V_{\textrm{mode}}}$ is the maximum spectrally-resolved SE rate enhancement~\cite{santori_single-photon_2010} and $\xi=\left(\frac{\vert\boldsymbol{\upmu}\cdot \textbf{E}\vert}{\vert\boldsymbol{\upmu}\vert\vert \textbf{E}_{\textrm{max}}\vert}\right)^2$ quantifies the angular and spatial overlap between the dipole moment ($\boldsymbol{\upmu}$) and the cavity mode electric field~($\textbf{E}$).

In contrast to atoms, quantum dots, and defect centers with narrow emission lines on the order of the cavity linewidth, the NV$^{-}$ emission has two major contributions, the narrow zero phonon line (ZPL) emission around 637~nm and a broad phonon sideband emission with a full-width at half maximum (FWHM) of about 100nm. The ratio between the two emission bands is described by the Debye-Waller factor $DW$ which is only about 3~\% for the NV$^{-}$, hence only a few percent ~\cite{zhao_suppression_2012} of the overall photoluminescence (PL) are emitted into the ZPL. Therefore, one has to differentiate between the overall Purcell enhancement $F$ and the spectrally-resolved spontaneous emission (SE) rate enhancement $F_{\textrm{ZPL}}=F/DW$ around the ZPL. 
\\

\subsection{Whispering-Gallery-Mode Resonators}
In an early demonstration of a whispering-gallery-mode resonator, diamond microdisks were fabricated into 
nanocrystalline diamond via FIB milling. Resonant modes with $Q$-factors of about 100 were observed near the NV$^{-}$ ZPL around 637~nm via detection of photoluminescence and near 1550~nm via evanescent fiber coupling~\cite{wang_observation_2007}. Suspended single crystal diamond microdisks were fabricated by implantation of 180~keV energy boron ions to create subsurface damage and homoepitaxial diamond overgrowth was applied for required microdisk thickness. The ion-damaged layer was selectively removed by electrochemical etching and the disks were patterned via ICP-RIE~\cite{wang_fabrication_2007}. The first resonant enhancement of the NV$^{-}$ ZPL was also realized with a single crystal diamond resonator that was patterned via EBL and oxygen RIE. A 10-fold enhancement was demonstrated, marking an important step of controlling the NV$^{-}$ emission via coupling to optical resonators~\cite{faraon_resonant_2011}. Further work with whispering gallery resonators coupled to waveguides will be discussed in the context of hybrid photonic systems in Section~\ref{sec:Hybrid}.

\begin{figure}[htbp]
\centering
\fbox{\includegraphics[width=\linewidth]{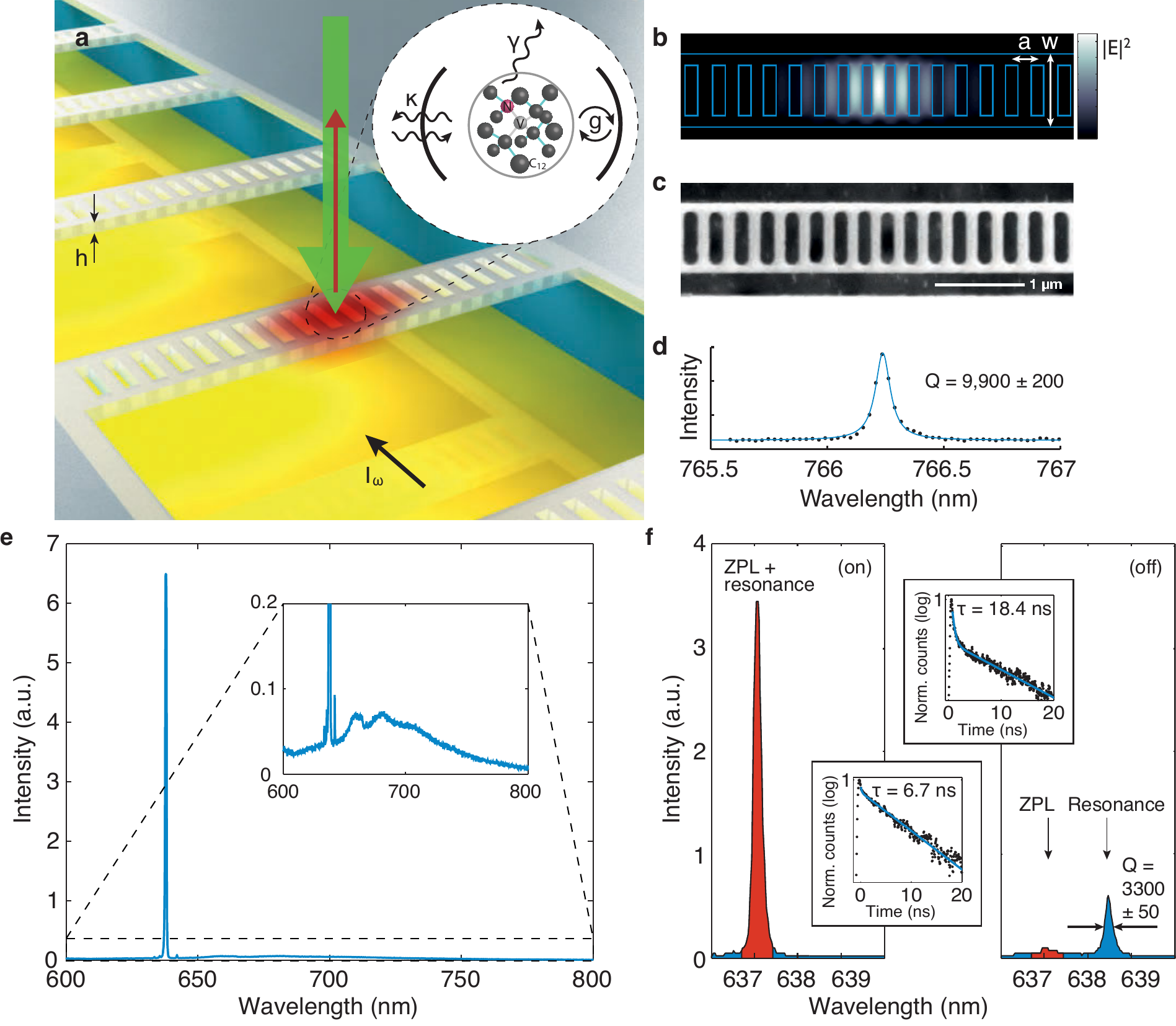}}
\caption{ a) The diamond photonic crystal (PhC) cavities are integrated on a Si substrate with metallic
striplines for coherent spin control and optically addressed using a confocal set up with 532nm continuous-wave excitation and photoluminescence collected $>630$~nm. The inset shows the nitrogen-vacancy (NV)-nanocavity system with g the NV$^{-}$-nanocavity Rabi frequency, $\gamma$ the NV$^{-}$ natural spontaneous emission (SE) decay rate and $\kappa$ the cavity intensity decay rate. The NV$^{-}$ consists of a substitutional nitrogen atom adjacent to a vacancy in the diamond lattice. $I_0$ denotes the current through the stripline, and h the membrane thickness. 
b) Simulated electric field intensity for the optimized fundamental cavity mode. The PhC has a width W and a lattice constant varying from 0.9a at the centre to a $\sim 220$~nm over five periods. c) Scanning electron
micrograph of a representative cavity structure. The scale bar represents 1 mm. d) Measured cavity resonance (dots) with a quality factor $\textrm{Q}=9,900 \pm 200$
from a Lorentzian fit (blue line).
e) System B at maximum Purcell enhancement. The inset shows a close-up of the spectrum. The ZPL transitions of four individual NV$^{-}$s (including the cavity-coupled ZPL)
are visible, each with a different strain-induced spectral position. The accumulated phonon sidebands of these NV$^{-}$s are also apparent. f) High resolution spectra of system B in cavity-coupled and uncoupled cases. The insets show the lifetime measurements corresponding to $\tau_{on} \sim 6.7$~ns and $\tau_{off} \sim 18.4$~ns. Reprinted with permission from Reference~\cite{li_coherent_2015}. \textcopyright 2015 Nature Publishing Group.}
\label{fig:thinfilmPhC}
\end{figure}

\subsection{Thin Film Photonic Crystal Cavities}
The first fabrication and optical characterization of photonic crystal cavities were demonstrated with nanocrystalline diamond, and fundamental cavity modes near the NV$^{-}$ ZPL with Q-factors up to 585 were observed~\cite{wang_fabrication_2007}. 1D-nanobeam photonic crystal cavities with theoretical Q-factors of up to 10$^6$ were introduced and fabricated via two different FIB milling methods~\cite{babinec_design_2011}. The first demonstration of coupling a single defect center to a PhC cavity was demonstrated by FIB milling of single crystal diamond for an L7~cavity design and a SiV$^{-}$ center with fluorescence intensity enhancement by a factor of 2.8~\cite{riedrich-moller_one-_2012}. The demonstration of the 70-fold enhancement of the ZPL transition rate of a cavity-coupled NV$^{-}$ marked an important step for cavity QED with defect centers in diamond, realized with a photonic crystal cavity fabricated in monocrystalline diamond using standard semiconductor fabrication techniques. The coupled NV$^{-}$ had a single-scan linewidth of a few GHz, determined with photoluminescence excitation measurements ~\cite{faraon_coupling_2012}. By coupling a single NV$^{-}$ to a waveguide based 1D-nanobeam photonic crystal cavity with Q-factors up to 6000, enhancement of the NV$^{-}$ ZPL fluorescence by a factor of $\sim 7$ was demonstrated~\cite{hausmann_coupling_2013}. Such waveguide based 1D-cavities enable the direct integration into a photonic architecture and are therefore interesting for efficient coupling and transmission experiments. A 1D-nanocavity fabricated by transferred hard mask lithography~\cite{li_nanofabrication_2015} and oxygen RIE (Fig.~\ref{fig:thinfilmPhC}) enabled the demonstration of Q-factors approaching 10,000, enhancement of the ZPL transition rate of $\sim 62$, and a beta-factor $\beta = 0.54$, indicating operation in the strong Purcell regime~\cite{li_coherent_2015}. Furthermore, electron spin manipulation was realized for the first time for cavity-coupled NV$^{-}$s with coherence times exceeding 200~$\upmu$s with on-chip microwave striplines for efficient spin control, providing a long-lived quantum system~\cite{li_coherent_2015}. This spin-photon interface experimentally validates the promise of long spin coherence NV$^{-}$-cavity systems for scalable quantum repeaters and quantum networks.

\subsection{Photonic Crystal Cavities in Bulk Diamond}
Due to the experimental difficulties of creating large scale, high-quality, membranes of uniform and controllable thickness, many groups have begun to explore the fabrication of photonic crystal cavities in bulk diamond. First implementations focused on creating membranes through ion damage of the diamond layer, and subsequent etching using FIB milling, as introduced in Sec.~\ref{sec:sec:ThinFilm}. This enabled photonic crystals etched from bulk diamond with Q$\sim500$ near the NV$^{-}$ ZPL~\cite{bayn_single_2010}. However, as discussed previously the lattice damage in this method is currently high and will most likely hinder the spin and spectral properties of defect centers. 

Three step tilted FIB milling, in which the stage is tilted with respect to the ion milling beam in two directions to achieve an undercut, and a last un-tilted mill step is used to etch the photonic crystal holes enabled nanobeam cavities which were separated from the bulk~\cite{bayn_triangular_2011,bayn_processing_2011}. This technology enabled cavities with Q's of a few hundred, matching theoretical predictions across multiple modes~\cite{bayn_triangular_2011}. Simulations also showed that the triangular geometry that results from the tilted FIB etching can support high Qs ($> 10^6$)~\cite{bayn_triangular_2011}. 

While FIB processing of diamond has enabled cavities in membrane and bulk diamond, it is an inherently low output process, as it is a serial etching process. The development of angled RIE etching with an angled cage allowed the extension of RIE etching techniques to triangular nanobeam cavities in bulk diamond. With this technique, cavities were realized with resonances near the NV$^{-}$ ZPL of a few thousand~\cite{bayn_fabrication_2014,burek_high_2014}. High Q cavities (loaded Q$>180,000$) have also been demonstrated in the infrared~\cite{burek_high_2014}. 

\section{All-Diamond Photonic Systems}
\label{sec:AllDiamond}
In contrast to diamond stand-alone devices, we will discuss diamond photonic systems comprising more than one optical element. This can, for example, be a ring-resonator coupled to waveguide. We make the distinction from stand-alone devices, as the extension of photonic systems can lead to complex architectures that will enable on-chip functionalities such as generation, entanglement, routing, and gating.

Two approaches have been demonstrated for the fabrication of all-diamond integrated photonic architectures. The first one is based on photonic elements etched into diamond-on-insulator or free-standing diamond thin-films. The second approach is based on fabricating monolithic, suspended 3D structures into bulk diamond samples. For more details on the patterning methods, please refer to Sec.~\ref{sec:DiamondPatterning}.

\subsection{Photonic Systems in Diamond Thin Films}
Silicon-on-insulator fabrication technology has enabled many high quality photonic structures by providing high index contrast and a stable platform. Diamond films suspended over air or placed on SiO$_2$ substrates provides similar index contrast and stability, and photonic device designs have been shown to translate with ease. The main disadvantage is the requirement of  large, single-crystal diamond films with uniform thickness, which are difficult to fabricate as discussed in Sec.~\ref{sec:sec:ThinFilm}.
Despite this challenge, first proof-of-principle experiments have been demonstrated. Haussmann et al. realized a nanophotonic network in a single crystal diamond film by integrating a high-Q ring resonator ($\textrm{Q} \sim 12600$) with an optical waveguide containing grating in- and outcouplers. A single NV$^{-}$ center inside the ring-resonator was coupled to its mode and single photon generation and routing was demonstrated with an overall photon extraction efficiency of about 10\%~\cite{hausmann_integrated_2012}. In a similar system, Faraon et al. showed strong enhancement of the zero-phonon line of NV$^{-}$ centers coupled to the ring resonance~\cite{faraon_quantum_2013}. By replacing grating couplers with polymer spot-size converters at the end of the diamond waveguides, off-chip fiber coupling as low as 1~dB/facet were demonstrated for wavelengths around 1550~nm. The integrated race-track resonators had quality factors up to $\sim 250,000$ and signatures of nonlinear effects were observed~\cite{hausmann_integrated_2013}. By coupling a second waveguide to a ring-resonator and by locally tuning the temperature of the diamond waveguides an optical-thermal switch was realized with switching efficiencies of 31\% at the drop and 73\% at the through port~
\cite{huang_microring_2013}.

\subsection{Photonic Systems in Bulk Diamond}
Initial attempts to fabricate photonic components in bulk diamond used a combination of patterning techniques (ion-induced damage for structure undercut, RIE for large pattern transfer, and FIB for local pattern transfer)~\cite{hiscocks_diamond_2008}. Two mode ridge waveguides in type-1b single crystal diamond were produced, though the damage caused by the ion implantation suggests that this technique cannot be adopted for quantum technologies. 

On the other hand, triangular RIE etching (as introduced in Sec.~\ref{sec:AngEtching}) is well suited to pattern photonic systems into bulk diamond directly. Free-standing components of a photonic integrated circuit, including optical waveguides and photonic crystal and microdisk cavities have been fabricated in a proof-of-principle demonstration~\cite{burek_free-standing_2012}. It was later shown in simulation that `s-bend' structures can be used in conjunction with the triangular RIE etching to add low-loss connection points to the bulk, and thus extend the length of the waveguides~\cite{bayn_fabrication_2014}. More work will have to be done in this area to increase the structural stability, operation at visible wavelengths, and fabrication yield of these bulk photonic integrated circuits. 

\section{Hybrid Photonic Systems}
\label{sec:Hybrid}

Hybrid photonic systems combine and exploit the advantages of multiple systems to achieve more functionality than any single isolated system. Random assembly, self assembly, bottom up fabrication, and manual assembly have all been used to access plasmonic and photonic regimes that would otherwise be inaccessible~\cite{benson_assembly_2011}. Here, we will review the integration of diamond with other semiconductor material systems to gain access to high quality cavity and photonic integrated circuit systems that are currently difficult to achieve with high-yield in diamond. Integration into cavities has allowed for emission enhancement of defects in un-patterned diamond slabs and nanodiamonds~\cite{santori_nanophotonics_2010}, and integration into waveguides has allowed for high collection efficiency and on-chip routing of emitted light.

\subsection{Cavity Systems}
\label{sec:HybCav}

\begin{figure}[htbp]
\centering
\fbox{\includegraphics[width=\linewidth]{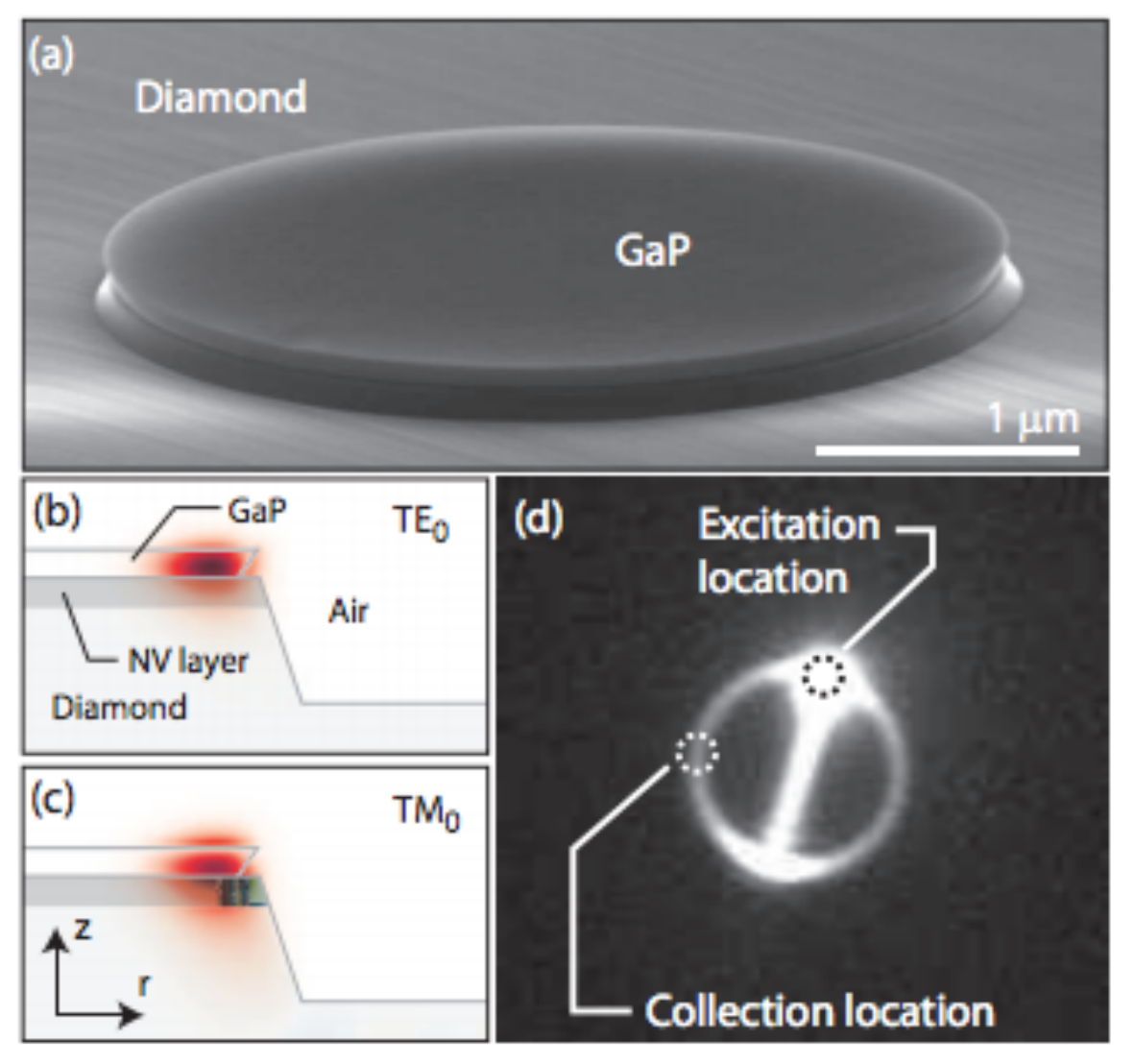}}
\caption{a) Scanning electron microscope (SEM) image of a hybrid
GaP-diamond microdisk. (b,c) FDTD simulated field profiles ($E_r(r,z)$ and $E_z(r,z)$, respectively), of the TE$_0^m$ and TM$_0^m$ modes.
d) Widefield CCD image of photoluminescence from a hybrid microdisk.
Reprinted with permission from Reference~\cite{barclay_chip-based_2009}. \textcopyright 2015 AIP Publishing LLC.
}
\label{fig:GaP}
\end{figure}

Due to its large bandgap, low intrinsic fluorescence, and ease of fabrication, Gallium Phosphide (GaP) has been used to enhance single defect emission in diamond. The float down of pre-patterned GaP microdisks onto an unpatterned bulk diamond allowed GaP to be used as both an etch mask and a higher index guiding material~\cite{barclay_chip-based_2009}. These high quality hybrid cavities supported whispering gallery mode resonances with $Q>25000$ and loaded Q factor of 3,800~\cite{thomas_waveguide-integrated_2014}. The structure and the mode profile can be seen in Fig.~\ref{fig:GaP}. However, the placement of NV$^{-}$s with respect to the optical mode is random, and single NV$^{-}$ enhancement was not shown. The same float down method was used to pattern ring resonators ($Q\sim3000-6800$) on a diamond sample with a lower density of NV$^{-}$ centers~\cite{barclay_hybrid_2011,fu_low-temperature_2011}. Tuning the cavity resonance at 10\,K, and measuring the cavity-coupled emission through a tapered fiber showed enhancement of multiple single NV$^{-}$ ZPLs. However, the Purcell effect was limited to $F\sim3.5$, again due to placement with respect to the mode maximum, as well as the large volume of the resonator.

Silica whispering gallery mode resonators (WGMs) have also been coupled to diamond structures with NV$^{-}$ centers in order to exploit the ultra-high quality factors possible with WGMs, although the mode volumes are high. 
A deterministic coupling approach in which silica microspheres are brought into contact with integrated 200\,nm diamond nanopillars with nanometer precision allows the preservation of NV$^{-}$ bulk properties while maintaining high quality factors ($Q > 10^6$) for the composite system~\cite{larsson_composite_2009}.

\subsection{Fiber Coupled DBR Cavity Systems}
High quality factor resonators can also be achieved with distributed Bragg reflector (DBR) cavities. Micro fabricated mirrors can facilitate high finesse, small mode volume cavities which are tunable post-fabrication~\cite{trupke_microfabricated_2005,hunger_laser_2012,muller_ultrahigh-finesse_2010,dolan_femtoliter_2010}. Integrating these DBR mirrors into a fiber-based system maintains the cavity properties, while achieving high coupling into a useable single fiber mode~\cite{steinmetz_stable_2006}. The coupling of a single NV$^{-}$ center in a nanodiamond to the field maximum of a tunable high finesse DBR cavity ($F=3500$ at $640\,$nm) via lateral positing allowed study of the phonon-assisted transitions of the NV$^{-}$~\cite{albrecht_coupling_2013}. A similar setup with a fiber-based DBR cavity elucidated the full scaling laws of Purcell enhancement for the NV$^{-}$ emission spectrum~\cite{kaupp_scaling_2013}. Both setups are projected to reach the strong Purcell regime at cold temperatures when the coupling between the NV$^{-}$ ZPL and the cavity field is larger. A fiber-based DBR cavity in which both input and output are coupled directly to fiber modes has increased the spectral photon rate density by orders of magnitude~\cite{albrecht_narrow-band_2014}, an important step for quantum information processing. These fiber-based DBR cavities have also been engineered to maintain high finesse and quality factors ($F=17000$ $Q\sim10^6$) even while containing thick diamond membranes ($>10\,\upmu$m) which can contain spectrally stable NV$^{-}$ centers with long coherence times, unlike nanodiamonds~\cite{janitz_fabry-perot_2015}.\\

\subsection{Photonic Circuit Integration}
Hybrid systems have also been used to enhance the coupling of the NV$^{-}$ emission into traveling-wave modes for enhanced detection rates, and collection into photonic modes that can be then manipulated and interfered to create larger networks. Early work has concentrated on the hybrid diamond-GaP systems discussed in Section~\ref{sec:HybCav}, demonstrating the evanescent coupling of NV$^{-}$ center emission to GaP multimode waveguides which suffered from high loss and fluorescence~\cite{fu_coupling_2008}. Theoretical work demonstrated the possibility of single-mode operation with better coupling between the NV$^{-}$ emission and the GaP waveguide mode, along with a scheme for coupling NV$^{-}$ centers more than 50nm from the center to high Q nanobeam cavities in the GaP layer~\cite{barclay_hybrid_2009}. Recent work on GaP-diamond hybrid systems has shown the waveguide-coupling of single NV$^{-}$ zero phonon line emission into disk resonators with estimated high zero-phonon line emission rates into one direction of the waveguide~\cite{gould_large-scale_2015}. This approach takes advantage of well-established thin film growth of and patterning methods of GaP, and enables the use of mainly unpatterned diamond to preserve the defect properties. However, this approach is limited by the reduced coupling of the defects to the waveguide mode. While this is mediated by the addition of a cavity as theoretically~\cite{barclay_hybrid_2009} and experimentally~\cite{gould_large-scale_2015} demonstrated, there exists no proposed way to locate the defect in the cavity mode maximum. 
\begin{figure}[htbp]
\centering
\fbox{\includegraphics[width=\linewidth]{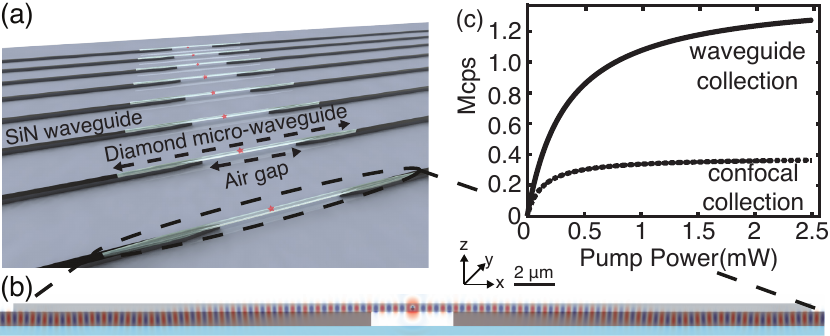}}
\caption{a) Sketch of a SiN PIC with multiple quantum nodes. b) FDTD simulation ($E_x$ field) showing the adiabatic-like mode transfer from a single-mode diamond waveguide into a single-mode. c) Saturation measurements acquired on the same emitter with confocal (dashed line) and waveguide (solid line) collection after background subtraction and correction for measured collection losses. About four times more emission is coupled into the waveguide compares to the high numerical aperture objective. Reprinted with permission from Reference~\cite{mouradian_scalable_2015}. \textcopyright 2015 APS.}
\label{fig:SiN}
\end{figure}

One approach to overcome this limitation is to fabricate diamond single-mode waveguides with NV$^{-}$s directly at the mode maximum. Such a waveguide can then be coupled with almost unity coupling efficiency to a pre-fabricated SiN photonic waveguide architecture with a suspended coupling region as shown in Fig.~\ref{fig:SiN}. For a diamond waveguide with a 200~x~200\,nm$^2$ cross section, it was determined that a dipole oriented perpendicularly to the propagation direction will couple $83\%$ of the emission to the single optical waveguide mode. Moreover, with the optimized tapering of both the diamond waveguide and the waveguide in the underlying photonic circuitry, up to $96\%$ of the light in the diamond waveguide will be coupled to the single mode SiN waveguide. In experiment, the hybrid structure shows that 3.5 times more photons emitted by the NV$^{-}$ are collected into one direction of the waveguide than into a free space 0.95\,NA objective, even with non-optimized diamond tapering regions~\cite{mouradian_scalable_2015}. A similar hybrid approach can also be used to efficiently couple light from single emitters in diamond to single-mode silica fibers. As in the photonic integrated circuit, tapers enable an adiabatic mode transfer between the single mode diamond waveguide and guided mode in the tapered silica fiber, theoretically enabling unity power transfer from diamond to fiber waveguide. Experimentally, an overall collection efficiency between 16\% and 37\% into a single optical mode was demonstrated, with a single photon count rate of more than 700\,kcps in saturation~\cite{patel_efficient_2016}. 

\subsection{Towards Complex Photonic Architectures}
Another important advantage of a hybrid bottom-up approach is that it enables the building of large-scale systems with almost unity probability, overcoming the stochastic defect center creation process with inherently low yield of high-quality quantum nodes. Pre-selection of the best diamond waveguides from an array of fabricated waveguides guarantees that every node in the final integrated network will contain a single defect with the desired spectral and spin properties, as well as being well coupled to the optical mode. This enables a linear scaling in the number of fabrication attempts necessary to create a quantum network with the desired number of nodes~\cite{mouradian_scalable_2015}.

To increase the detection efficiency of single photon emitted by defects in diamond, efforts are underway to fabricate superconducting nanowire single photon detectors (SNSPDs) directly on diamond or to integrated them with hybrid waveguide architectures~\cite{sprengers_waveguide_2011,najafi_-chip_2015}. Niobium Titanium Nitritide (NbTN) SNSPDs have been fabricated on single-crystal diamond substrates and have been shown to have good detection properties~\cite{atikian_superconducting_2014}. Niobium nitride (NbN) SNSPDs fabricated directly on waveguides in polycrystalline thin-film diamond grown on oxide show high detection efficiencies up to 66\% at 1550\,nm combined with low dark count rates, and timing resolution of 190\,ps~\cite{rath_superconducting_2015}. \\

\section{Conclusion and Remaining Challenges}
As discussed in this review, recent advances in diamond synthesis and fabrication have enabled high-quality nano- and microphotonic devices for increased photon collection and tailored light-matter interaction. Despite this progress, there are still fundamental challenges to be overcome on the way towards more complex QI implementations based on these diamond photonic nanostructures. High fidelity QI applications require long spin coherence times and lifetime limited emission linewidths~\cite{kimble_quantum_2008,duan_scalable_2004,duan_long-distance_2001}. As noted in the review, many diamond growth and fabrication schemes have been tailored to bolster spin coherence times. However, more development must be done to produce nanostructures with both high intrinsic quality (high Q values and low mode volumes) and high defect quality (long spin coherence times and lifetime limited emission linewidths). In particular the latter has up to date been limited to a factor of about 30 of the fluorescence lifetime-limit \cite{mouradian_scalable_2015}. It is commonly believed that both crystal lattice defects induced by dry etching and increased surface area near defects due to nanofabrication lead to the degradation of defect properties. Therefore, suitable nanofabrication technology and proper surface termination methods must be developed. Furthermore, the realization of large-scale quantum photonic systems  will depend on \textit{scalable fabrication} techniques and on \textit{tunability} -- Stark shift control to bring multiple defect transitions (e.g. the NV$^-$ ZPL) to the same frequency~\cite{tamarat_stark_2006}, and cavity tuning~\cite{chew_enhanced_2011} to match the resonance wavelength with the defect transition frequency without degrading the cavity Q. Finally, for high fidelity information processing, error correction is required. To establish error-corrected logical qubit nodes in a large quantum photonic processor \cite{blok_towards_2015}, advances must be made in the high yield creation of coupled defect centers \cite{dolde_high-fidelity_2014} as well as control over the surrounding nuclear environment as resource to potentially store quantum states longer than an individual quantum memory is able to \cite{waldherr_quantum_2014,taminiau_universal_2014,cramer_repeated_2015}. The progress presented in this review proves the viability of diamond based nanophotonic systems in QI and sensing, and further progress will enable enhanced sensing as well as scalable solid state quantum networks.

\begin{acknowledgments}
Fabrication and experiments were supported in part by the Air Force Office of Scientific Research PECASE (AFOSR Grant No. FA9550-11-1-0014), the AFOSR Quantum Memories MURI, and the U.S. Army Research Laboratory (ARL) Center for Distributed Quantum Information (CDQI). Research was carried out in part at the Center for Functional Nanomaterials, Brookhaven National Laboratory, which is supported by the U.S. Department of Energy, Office of Basic Energy Sciences, under Contract No. DE-SC0012704. T.S. was supported in part by the CDQI. E.H.C. was supported by the NASA Office of the Chief Technologist's Space Technology Research Fellowship. M.E.T. was supported in part by the AFOSR Quantum Memories MURI. M.E.T. was supported by the NSF IGERT program Interdisciplinary Quantum Information Science and Engineering (iQuISE). S.M. is supported by the AFOSR PECASE.
\end{acknowledgments}


\end{document}